\documentclass[pra,nofootinbib,superscriptaddress,showpacs,showkeys]{revtex4}

\usepackage{graphicx,epsfig}
\usepackage{amsfonts,amsmath,amssymb,amsthm,amscd}
\usepackage[T2A]{fontenc}
\usepackage{color}

\begin{document}

\title{Electromagnetic cascade in high energy electron, positron, and photon interactions with intense laser pulses}

\author{S. S. Bulanov}
\affiliation{University of California, Berkeley, California 94720, USA}

\author{C. B. Schroeder}
\affiliation{Lawrence Berkeley National Laboratory, Berkeley, California 94720, USA}

\author{E. Esarey}
\affiliation{Lawrence Berkeley National Laboratory, Berkeley, California 94720, USA}

\author{W. P. Leemans}
\affiliation{University of California, Berkeley, California 94720, USA}
\affiliation{Lawrence Berkeley National Laboratory, Berkeley, California 94720, USA}

\begin{abstract}
The interaction of high energy electrons, positrons, and photons with intense laser pulses is studied in head-on collision geometry. It is shown that electrons and/or positrons undergo a cascade-type process involving multiple emissions of photons. These photons can consequently convert into electron-positron pairs. As a result charged particles quickly lose their energy developing an exponentially decaying energy distribution, which suppresses the emission of high energy photons, thus reducing the number of electron-positron pairs being generated. Therefore, this type of interaction suppresses the development of the  electromagnetic avalanche-type discharge, i.e., the exponential growth of the number of electrons, positrons, and photons does not occur in the course of interaction. The suppression will occur when 3D effects can be neglected in the transverse particle orbits, i.e., for sufficiently broad laser pulses with intensities that are not too extreme. The final distributions of electrons, positrons, and photons are calculated for the case of a high energy e-beam interacting with a counter-streaming, short intense laser pulse.  The energy loss of the e-beam, which requires a self-consistent quantum description, plays an important role in this process, as well as provides a clear experimental observable for the transition from the classical to quantum regime of interaction.           
\end{abstract}

\maketitle

\section{Introduction}

The processes typical for High Intensity Particle Physics \cite{BulanovAAC12}, {\it i.e.}, the interactions of charged particles with strong electromagnetic fields \cite{reviews}, have attracted considerable interest recently. This interest is due to the rapid growth of the maximum achievable laser intensity at many existing, constructed, and projected experimental facilities. Some of these processes, previously believed to be of theoretical interest only, are now becoming experimentally accessible. 

\begin{figure}[tbp]
\begin{center}
\includegraphics[width=100mm]{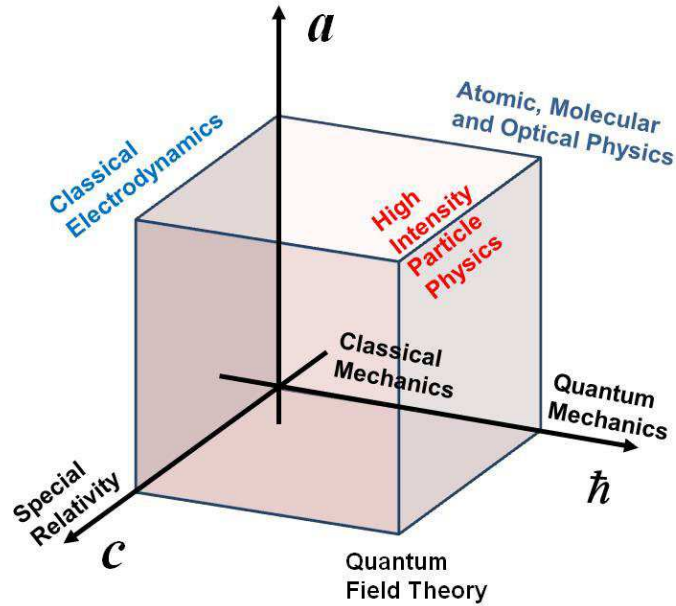}
\end{center}
\caption{(Color online) The cube of theories: three axis correspond to relativistic ($c$), quantum ($\hbar$), and high intensity effects ($a$); the vertices of the cube - $(0,0,0)$ is classical mechanics, $(c,0,0)$ is Special Relativity, $(0,\hbar,0)$ is Quantum Mechanics, $(c,\hbar,0)$ is Quantum Field Theory, $(c,0,a)$ is the classical Electrodynamics, $(0,\hbar,a)$ is atomic, molecular and optical physics, and $(c,\hbar,a)$ is the High Intensity Particle Physics.}\label{cube}
\end{figure}

High intensity electromagnetic (EM) fields significantly modify the interactions of particles and EM fields, giving rise to the phenomena that are not encountered either in classical or perturbative quantum theory of these interactions. One can imagine a cube of theories (analogous to the cube mentioned in Ref. \cite{okun}), which is located along three orthogonal axes marked by $c$,  $\hbar$, and $a$ (see Fig. \ref{cube}). These three axis correspond to relativistic, quantum, and high intensity effects. Each vertex of the cube corresponds to a physical theory: $(0,0,0)$ is non-relativistic mechanics, $(c,0,0)$ is Special Relativity, and $(0,\hbar,0)$ is Quantum Mechanics. The theory that has both quantum and relativistic effects included is the Quantum Field Theory, $(c,\hbar,0)$. The classical Electrodynamics corresponds to the vertex $(c,0,a)$ and atomic, molecular and optical physics to $(0,\hbar,a)$. If the high intensity effects are included in the framework of the Quantum Field Theory, then the corresponding vertex $(c,\hbar,a)$ corresponds to the High Intensity Particle Physics. Thus the high intensity EM fields add a new dimension to the processes occurring in Quantum Field Theory, significantly changing the physics of the interactions.   

The intensity of $2\times 10^{22}$ W/cm$^2$ was demonstrated several years ago \cite{Yanovsky} and the intensity of $10^{23}$ W/cm$^2$ can in principle be achieved at several PW-laser facilities, which are being built or are already operational. Several projected facilities are aiming to reach intensities of $10^{26}$ W/cm$^2$ \cite{ELI}. At laser intensities of $10^{23}$ W/cm$^2$ and above the electromagnetic (EM) radiation interaction with charged particles becomes highly dissipative, due to the efficient generation of high energy $\gamma$-rays \cite{reviews,radiation,NakamuraPRL2012,RidgersPRL2012, ThomasPRX2012}. These high energy photons can decay in the strong EM field producing $e^+e^-$ pairs, which in turn will lose their energy emitting photons. 

It is widely discussed in the literature \cite{BellKirk, FedotovPRL2010, BulanovPRL2010} whether these two processes occurring subsequently in the EM field of two colliding laser pulses will give rise to the EM avalanche-type discharge at intensities of about $10^{25}$ W/cm$^2$, {\it i.e.}, the exponential growth of the number of electrons, positrons, and photons due to the fact that the charged particles are being constantly accelerated by the EM field. This phenomenon would significantly change the properties of the laser pulse interaction with the charged particles, and the most evident consequence will be the scattering of the laser radiation at the generated $e^+e^-\gamma$ plasma \cite{BulanovPRL2010}. For example, this scattering might limit the maximum attainable laser intensity \cite{FedotovPRL2010, BulanovPRL2010}, which may occur much earlier than the laser pulse depletion due to pair production from vacuum, discussed in Refs. \cite{pair production}.  

It is convenient to parametrize the charged particle interaction with the EM field in terms of the dimensionless amplitude of the EM field vector-potential:
\begin{equation}
a=\frac{eE}{m\omega_0 c},
\end{equation} 
where $e$ and $m$ are electron charge and mass respectively, $c$ is the speed of light, $\omega_0$ and $E$ are the frequency and strength of the EM field respectively. The parameter $a$ has a meaning of an electron energy gain over a distance of one wavelength in units of its rest energy, $mc^2$. The value $a=1$ marks the onset of the relativistic regime in the charged particle interaction with the EM field. The possibility of new particle production by the EM field is connected with a field strength that is able to perform $mc^2$ work over the electron Compton length, $\lambda_C=\hbar/mc=3.86\times10^{-11}$ cm, \textit{i.e.}, $eE_S\lambda_C=mc^2$. This field, 
\begin{equation}
E_S=\frac{m^2 c^3}{e\hbar}=1.32\times 10^{16}~~\text{V/cm},
\end{equation}
is usually referred to as the Quantum Electrodynamics (QED) "critical" field \cite{Schwinger}. The subscript "S" in the definition of the "critical" field as well as in the subsequent definitions of the "critical" intensity and vector-potential stands for the fact that this field is usually referred to as a "Sauter-Schwinger" field. The dimensionless vector-potential corresponding to this field is 
\begin{equation}
a_S=\frac{mc^2}{\hbar\omega_0}.
\end{equation}
and has a meaning of a minimal number of photons that need to be absorbed from the field to produce a new particle. For a laser pulse with 1 $\mu$m wavelength $a_S=4.1\times 10^5$. 

The "critical" field of QED is unaccessible in the near future by laser technology, since the corresponding intensity is $I_S=4.65\times 10^{29}$ W/cm$^2$. Even the less demanding peak intensities that are required for scenarios of electron-positron pair production by single focused, two counter-propagating \cite{pair production} or multiple focused at one spot \cite{multiple pulses} laser pulses, which require peak intensity of $10^{25-28}$ W/cm$^2$, will not become available in the near term. However a high energy electron/positron or photon can experience such a field in collision with the high intensity laser pulse. Such interaction is characterized by the parameter $\chi_e$ for an electron/positron and $\chi_\gamma$ for a photon:
\begin{eqnarray}\label{chi}
\chi_e=\frac{e\hbar\sqrt{(F_{\mu\nu}p^\nu)^2}}{m^3c^4}=\frac{1}{E_S}\sqrt{\left(\gamma \mathbf{E}+\frac{\mathbf{p\times\mathbf{B}}}{mc}\right)^2-\left(\frac{\mathbf{p}\cdot\mathbf{E}}{mc}\right)^2}, \\
\chi_\gamma=\frac{e\hbar\sqrt{(F_{\mu\nu}k_\gamma^\nu)^2}}{m^3c^4}=\frac{1}{E_S}\sqrt{\left(\hbar\omega_\gamma \mathbf{E}+\frac{\mathbf{k_\gamma\times\mathbf{B}}}{mc}\right)^2-\left(\frac{\mathbf{k}_\gamma\cdot\mathbf{E}}{mc}\right)^2}.
\end{eqnarray}  
Here $p$ is the electron momentum, $\gamma=\sqrt{1+(p/mc)^2}$, and the photon momentum is $k_\gamma=(\hbar\omega_\gamma,\mathbf{k}_\gamma)$. The parameter $\chi_e$ has a meaning of the EM field strength normalized to the "critical" field strength in the reference frame where the electron is at rest. The parameters $a$ and $\chi_{e,\gamma}$ play an important role in calculating the probabilities of QED processes in strong EM fields: multiphoton Compton scattering ($e\rightarrow e\gamma$) \cite{BrownKibble,RitusJETP1964,Goldman} and multiphoton Breit-Wheeler process ($\gamma\rightarrow ee$) \cite{BreitWheeler,Reiss,RitusJETP1964,Yakovlev}. When $a\ll 1$ the interaction of an electron with an EM wave can be considered as an interaction with a single photon. The probabilities of interaction with two or more photons are negligible in this case. For $a\ge 1$ these probabilities become comparable with the one-photon interaction probability. Thus the process becomes multiphoton, \textit{i.e.}, it acquires a nonlinear dependence on the field amplitude. The parameters $\chi_{e,\gamma}$ govern the magnitude of quantum effects.  

For an electron interaction with a plane EM wave propagating along x-axis with $E=E(x-t)e_y$ and $B=E(x-t)e_z$: $\chi_e=(E/E_S)(\gamma-p_x/mc)$, and for a photon interaction with this wave: $\chi_\gamma=(E/E_S)(\hbar\omega-k_x c)/mc^2$. Here we used the conservation of the electron generalized momentum, which gives the component of the electron momentum parallel to the laser electric field. One can see that the quantum effects are maximized for an electron/positron or a photon counter-propagating with the EM wave. In the ultra-relativistic limit, $\gamma\gg 1$:
\begin{equation}
\chi_e^{\uparrow\downarrow}= 2\gamma\frac{E}{E_S},~~~\chi_\gamma^{\uparrow\downarrow}\simeq 2\frac{\hbar\omega}{mc^2}\frac{E}{E_S}
\end{equation}   
Here $\uparrow\downarrow$ denotes the counter-propagating electron/positron or photon and the EM wave. If an electron/positron or a photon co-propagates with the EM wave, then in the former case the parameter $\chi_e$ is reduced and in the later case the parameter $\chi_\gamma$ is equal to zero:
\begin{equation}
\chi_e^{\uparrow\uparrow}\simeq(2\gamma)^{-1}\frac{E}{E_S},~~~\chi_\gamma^{\uparrow\uparrow}=0,
\end{equation}
where $\uparrow\uparrow$ denotes the co-propagating electron/positron or photon and the EM wave. The vanishing of the photon parameter $\chi^{\uparrow\uparrow}_\gamma=0$ is connected with the fact that co-propagating massless particles do not interact, since the terms corresponding to the photon-photon scattering in the EM field Lagrangian vanish for this interaction geometry (on the possibility of measuring the photon-photon scattering for eV-keV photons see Ref. \cite{koga1}). Analogous effects in gravitational interaction of massless particles were mentioned by Zee in Ref. \cite{Zee}. Since we are interested in exploring the effects of High Intensity Particle Physics, which optimally requires $\chi_e\sim 1$, the configuration of a counter-propagating electron beam and a laser pulse is most beneficial for studying these effects.

The configuration of a counter-propagating 46 GeV electron beam and a $10^{18}$ W/cm$^2$ laser pulse was used in the E144 experiment at SLAC \cite{Bula1996}. The resulting EM field strength in the electron rest frame was a quarter of the QED critical field. However, the relatively low laser intensity resulted in long mean free path of electrons and photons with respect to either radiation or pair production and thus led to a small number of events observed. By contrast, today the achievable peak laser intensity is of the order of $10^{22}$ W/cm$^2$, GeV electron beams are routinely produced by laser-plasma accelerators \cite{Leemans,Esarey_RMP}, and there are projects to achieve 10 GeV electron beams in the near future \cite{BELLA}. Thus combining a 10 GeV electron beam with a $10^{22}$ W/cm$^2$ laser pulse will bring us much further into the experimentally uncharted domain of High Intensity Particle Physics \cite{BulanovPRL2010}. 

As it was shown in \cite{BulanovPRL2010,ThomasPRX2012} the effects of nonlinear QED begin to dominate when the emitted photon carries away an energy on the order of the electron energy, $\hbar\omega_\gamma\approx\gamma m c^2$, where $\omega_\gamma$ is the frequency of the emitted photon. For a head-on collision  of an electron and a laser pulse, the emitted photon energy is $\hbar\omega_\gamma\approx 1.2\hbar\omega_0 a\gamma^2$ and the number of emitted photons per laser pulse period is $N_\gamma\approx (3\pi/4)\alpha a$ \cite{BulanovPRL2010}, where $\alpha=1/137$ is the fine structure constant. Hence the effects of High Intensity Particle Physics should be taken into account for 
\begin{equation}\label{a_Q}
a>a_Q=(2/3)\alpha(1.2\epsilon_{rad}\gamma)^{-1},
\end{equation}
where $\epsilon_{rad}=4\pi r_e/3\lambda_0$, $r_e=2.8\times 10^{-13}$ cm is the classical electron radius, and $\lambda_0$ is the laser wavelength. Eq. (\ref{a_Q}) is analogous to two conditions, derived from the analysis of the probability of the photon emission by an electron, $\chi_e>1$ and $\alpha a>1$ \cite{diPiazzaPRL2010}. The first condition indicates that the recoil in each photon emission is important, and the second one indicates that the number of photons emitted incoherently per laser period can be larger than unity. For a 10 GeV electron beam and a laser with a wavelength of $0.8~\mu$m, $a_Q\approx 20$. This value of $a_Q$ is below the peak value of $a$, which was already demonstrated in the experiments by focusing the laser pulse to the intensity of $10^{22}$ W/cm$^2$, corresponding to $a\approx 70$. Thus it is possible to explore experimentally high intensity effects with present laser systems. Multiphoton Compton and multiphoton Breit-Wheeler processes have attracted a lot of attention recently for this reason. Especially the interaction of electrons, positrons, and photons with finite duration laser pulses, and effects that accompany it, were studied \cite{Harvey,Heinzl,diPiazzaPRL2010,Sokolov,Sokolov_kinetic, Mackenroth,Titov}.         

In this paper we study the interaction of an energetic electron beam with the intense laser pulse in the framework of High Intensity Particle Physics. We aim at the theoretical exploration of the interaction regimes characterized by high energy dissipation and associated processes. The individual processes of a photon emission by an electron and a photon decay into an electron-positron pair in strong constant crossed field show an enhancement in the production of high energy particles. However, in the interaction with the laser pulse, the electron beam undergoes a cascade-type process involving multiple emissions of photons and the consequent decay of these $\gamma$'s into electron-positron pairs, which can be described by the system of kinetic equations for the electron, positron and photon distribution functions, analogous to the approach of Refs. \cite{kinetic,Sokolov_kinetic,kinetic_Bell}. As a result the initial electron beam quickly loses its energy and develops a broad energy distribution with an exponentially decaying high energy tail. For this type of energy distribution the emission of high energy $\gamma$'s is suppressed, which leads to a reduction of $\gamma$'s decays into $e^+e^-$ pairs. Thus this type of interaction disfavors the EM avalanche-type discharge development. 

We also compare the approach based on the QED rates for $e\rightarrow\gamma e$ and $\gamma\rightarrow ee$ processes and corresponding kinetic equations with the solution of classical equations of motion with radiation friction force included. The importance of the multiphoton absorption for the electron dynamics in the strong EM pulse is discussed. 

The paper is organized as follows. In section 2 we consider the dependence of the probabilities of the $e\rightarrow e\gamma$ and $\gamma\rightarrow ee$ processes in constant crossed EM field on various parameters and study different limiting cases. In section 3 we present the results on the cascade formation and the comparison between the "quantum" and "classical" cases. We conclude in section 4.

\section{Multiphoton Compton and Breit-Wheeler processes in a constant crossed EM field}

\begin{figure}[tbp] 
\epsfxsize10cm\epsffile{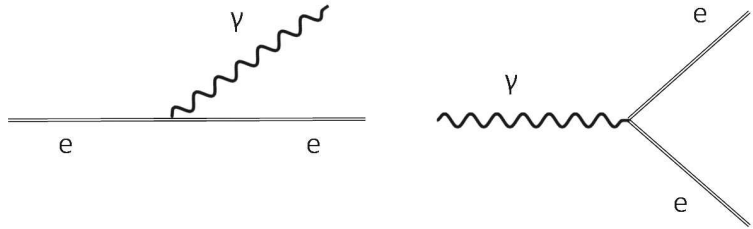}
\caption{The Feynman diagrams of the Compton ($e\rightarrow e\gamma$) and Breit-Wheeler ($\gamma\rightarrow ee$) processes. Double fermion lines indicate that the process occurs in external field.} \label{diagram}
\end{figure}

In what follows we consider the processes of a photon emission by an electron and a photon conversion into an electron-positron pair in a strong EM field (see Fig. \ref{diagram}):
\begin{equation}\label{compton}
e(p)\rightarrow e(p^\prime)\gamma(k^\prime)
\end{equation}
\begin{equation}\label{breit-wheeler}
\gamma(k)\rightarrow e(p^\prime)e(q^\prime),
\end{equation}
where $p$, $q$ and $k_\gamma$ are momenta of initial electron, positron and photon respectively and primed momenta refer to the same particle but in the final state. The electric and magnetic field strengths are denoted as $\mathbf{E}$ and $\mathbf{B}$ respectively, and the wave vector as $k_0$. In this paper a system of units is chosen such that  $c=1$ and $\hbar=1$. Based on the results of \cite{RitusJETP1964}, we study the properties of processes (\ref{compton}) and (\ref{breit-wheeler}) in the plane EM wave with regard to particle energies and field strength.

In what follows we introduce parameters $\chi_e=\chi_e(p)$, $\chi_e^\prime=\chi_e(p^\prime)$, $\chi_\gamma=\chi_\gamma(k_\gamma)$, and $\chi_\gamma^\prime=\chi_\gamma(k^\prime_\gamma)$, which describe the strength of the interaction of initial and final state electrons and photons with the EM field (for positrons we also use the notation $\chi_e=\chi_e(q)$, $\chi_e^\prime=\chi_e(q^\prime)$ and since these parameters for an electron and positron do not appear together it should not cause any confusion).  These parameters are not independent, they are connected by the energy-momentum conservation:
\begin{equation}\label{energy conservation}
sk_0+p=p^\prime+k_\gamma^\prime,
\end{equation}   
where $s$ is the number of photons absorbed from the EM field ($s\omega_0+p_0=p^\prime_0+\omega^\prime_\gamma$). If we multiply this equation by $k_0$ and then using the fact that $k_0^2=0$ we rewrite it in terms of the invariants to obtain 
\begin{equation}
\chi_e=\chi_e^\prime+\chi_\gamma^\prime.
\end{equation}

In what follows we consider a simplified 1D model of the electron interaction with the laser pulse. We assume that initially electron momentum is directed along the laser pulse propagation axis, but oppositely to the direction of the laser pulse, $\mathbf{p}=(-p_{x},0,0)$, and in the course of interaction the y- and z-components do not change, {\it i.e.}, we assume $p_0\gg 1$ and the dynamics of electrons and positrons is dominated by the longitudinal motion. This implies that a transverse quiver amplitude of the electron, $\sim \lambda_0 a/\gamma $, is much less than the laser spot size $r_0$. In this case $\chi_e=(E/E_S)(\gamma_e-p_x/m_e)$. It was pointed out in Ref. \cite{RitusJETP1964} that since the wavelength of the laser is much larger than the characteristic scale of the formation of the process, $a^{-1}$, the laser field can be regarded as constant. Moreover for an ultra-relativistic electron beam the EM field of the laser in the electron rest frame is very similar to the crossed EM field ($\mathbf{E}\perp \mathbf{B}$, $E=B$). Therefore we assume the approximation of a locally constant crossed EM field and consider the spectra and probabilities of multiphoton Compton and Breit-Wheeler processes in such a field. The differential probabilities of these processes can be written in the following form:
\begin{equation} \label{dPe}
dP^e=-\frac{\alpha}{\pi \lambda_C}\frac{m_e}{\gamma}F_+\left(z_e,y_e\right)d\epsilon_\gamma,
\end{equation}
\begin{equation} \label{dPgamma}
dP^\gamma=-\frac{\alpha}{\pi \lambda_C}\frac{m_e}{\omega_\gamma}F_-\left(z_\gamma,y_\gamma\right)d\epsilon_e,
\end{equation}
where $\epsilon_\gamma=k_\gamma^\prime/p_0$ and $\epsilon_e=p_0^\prime/k_\gamma$ are the emitted photon energy normalized to the initial electron energy and the produced electron energy normalized to the initial photon energy respectively. The function $F(z,y)$ is
\begin{equation}
F_\pm\left(z,y\right)=\frac{2}{3\mp 1}\left[\int\limits_z^\infty \Phi(z^\prime)dz^\prime\pm\frac{y}{z}\Phi^\prime(z)\right].
\end{equation}
Here $\Phi(z)$ and $\Phi^\prime(z)$ are the Airy function and its derivative, respectively, which can be expressed in terms of a modified Bessel function of a second kind: $\Phi(z)=3^{-1/2}\pi^{-1}z^{1/2}K_{1/3}(2z^{3/2}/3)$, and $\Phi^\prime(z)= -3^{-1/2}\pi^{-1}z K_{2/3}(2z^{3/2}/3)$ \cite{A&S}. 

\begin{figure}[tbp]
\epsfxsize10cm\epsffile{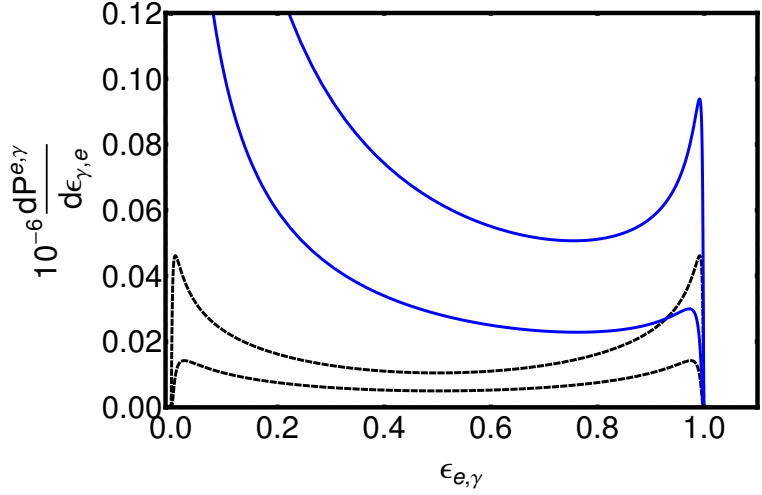}
\caption{(Color online) The spectra of photons from the $e\rightarrow e\gamma$ process (blue solid curves) and electrons from the $\gamma\rightarrow ee$ process (black dotted curves) in a constant crossed field for different values of the field intensity: $10^{24}$, $10^{25}$ W/cm$^2$. The energy of the initial electron and photon is 10 GeV.} \label{dP}
\end{figure}

Though expressions for the differential probabilities of the multiphoton Compton and Breit-Wheeler processes (\ref{dPe},\ref{dPgamma}) look similar, the fact that in one case the electron and photon are in the final state, whereas in the other case the electron and positron are produced, is accounted for in the explicit form of variables $z_{e,\gamma}$ and $y_{e,\gamma}$:
\begin{equation}
z_e=\left[\frac{\epsilon_\gamma}{(1-\epsilon_\gamma)\chi_e}\right]^{2/3},~~~y_e=1-\epsilon_\gamma+\frac{1}{1-\epsilon_\gamma}, 
\end{equation}
\begin{equation}
z_\gamma=\left[\frac{1}{\epsilon_e(1-\epsilon_e)\chi_\gamma}\right]^{2/3},~~~y_\gamma=\frac{1-\epsilon_e}{\epsilon_e}+\frac{\epsilon_e}{1-\epsilon_e}.
\end{equation}
If we compare two cases where (i) almost all the energy of an initial electron is transferred to an emitted photon, $\epsilon_\gamma\rightarrow 1$, and (ii) almost all energy of an initial photon is transferred to either final electron or a final positron, $\epsilon_e\rightarrow 1$, then the behavior of the variables $z_{e,\gamma}$ and $y_{e,\gamma}$ is the same. They all tend to infinity: $z_{e,\gamma}\rightarrow \infty$ and $y_{e,\gamma}\rightarrow \infty$. In the opposite situation we compare (i) the case where the emitted photon energy tends to zero, $\epsilon_{e}\rightarrow 0$, and (ii) the case where either final electron or final positron is produced almost at rest $\epsilon_{\gamma}\rightarrow 0$.  Here the behavior of the variables $z_{e,\gamma}$ and $y_{e,\gamma}$ is completely different. While $z_{e}$ tends to zero  and $y_{e}$ tends to 2, both $z_{\gamma}$ and $y_{\gamma}$ tend to infinity. This means that the differential probabilities for multiphoton Compton and Breit-Wheeler processes in the case of a maximum asymmetry in the momentum distribution among the final state particles should be almost similar. Moreover the fact that for both $\epsilon_e\rightarrow 1$ and $\epsilon_e\rightarrow 0$ the variables $z_{\gamma}$ and $y_{\gamma}$ tend to infinity indicates that the differential probability for Breit-Wheeler process is symmetric with respect to $\epsilon_e\rightarrow 1-\epsilon_e$. These properties can be seen from Fig. \ref{dP} where the dependencies of differential probabilities on $\epsilon_\gamma$ for multiphoton Compton and $\epsilon_e$ for multiphoton Breit-Wheeler are shown. 

\subsection{Differential probabilities of multiphoton Compton and Breit-Wheeler processes: $\epsilon_{\gamma,e}\rightarrow 1$}

In the limit $\epsilon_{e,\gamma}\rightarrow 1$ the integration in Eqs. (\ref{dPe},\ref{dPgamma}) can be carried out and the differential probabilities will take the following form:
\begin{equation} \label{dP^e_1}
dP^e=\frac{\alpha}{\pi \lambda_C}\frac{m_e}{\gamma}\frac{\chi_e^{1/2}}{\left(1-\epsilon_\gamma\right)^{1/2}}\exp\left[-\frac{2}{3}\frac{\epsilon_\gamma}{(1-\epsilon_\gamma)\chi_e}\right]d\epsilon_\gamma,
\end{equation}
\begin{equation}\label{dP^gamma_1}
dP^\gamma=-\frac{\alpha}{2\pi \lambda_C}\frac{m_e}{\omega_\gamma}\frac{\chi_\gamma^{1/2}}{\left(1-\epsilon_e\right)^{1/2}}\exp\left[-\frac{2}{3}\frac{1}{\epsilon_e(1-\epsilon_e)\chi_\gamma}\right]d\epsilon_e.
\end{equation}
The functions in Eqs. (\ref{dP^e_1},\ref{dP^gamma_1}) demonstrate almost identical behavior $\sim \delta\epsilon^{-1/2}\exp\left(-2/3\delta\epsilon\chi_{e,\gamma}\right)$, where $\delta\epsilon=1-\epsilon_{e,\gamma}$. For $\chi_{e,\gamma}\gg 1$ these functions have a maximum near $\epsilon_{e,\gamma}=1$:
\begin{equation}\label{max energy}
\epsilon_{e,\gamma}=1-\frac{4}{3\chi_{e,\gamma}},
\end{equation}
which corresponds to the enhancement of the high energy electrons/positrons or photons production in $\gamma\rightarrow ee$ and $e\rightarrow\gamma e$ processes respectively. In the case of $e\rightarrow\gamma e$ the requirement $\chi_e>4/3$ should be satisfied for the maximum in $dP^e/d\epsilon_\gamma$ to exist. In the case of $\gamma\rightarrow ee$ this requirement is $\chi_\gamma>8$. If either the final photon in multiphoton Compton process or the final electron/positron in multiphotin Breit-Wheeler process was produced with the energy given by Eq. (\ref{max energy}) then the other particle in the final state will have an energy of $4/3\chi_{e,\gamma}$ or $(4/3)(E_S/E)$, which does not depend on the energy of an initial state particle. Since the production of particles with $\epsilon_{e,\gamma}>1-4/3\chi_{e,\gamma}$ is exponentially suppressed (see Eqs. (\ref{dP^e_1},\ref{dP^gamma_1})), this result introduces a low energy cutoff for electrons and positrons:
\begin{equation}
p_{th}=\frac{4}{3}m_e\frac{E_S}{E},
\end{equation}   
which holds as long as the condition $\chi_{e,\gamma}\gg 1$ is satisfied.

\subsection{Differential probabilities of multiphoton Compton and Breit-Wheeler processes: $\epsilon_{\gamma,e}\rightarrow 0$}

In the opposite case of $\epsilon_{e,\gamma}\rightarrow 0$ the differential probabilities of $e\rightarrow\gamma e$ and $\gamma\rightarrow ee$ processes demonstrate different behavior. While $dP^\gamma/d\epsilon_e$ is symmetric with respect to $\epsilon_e\rightarrow 1-\epsilon_e$, $dP^e/d\epsilon_\gamma$ is not with respect to $\epsilon_\gamma\rightarrow 1-\epsilon_\gamma$, due to the fact that there is no charge symmetry in the final state. The differential probability of emitting a low energy photon goes to infinity as $\epsilon_\gamma^{-2/3}$:
\begin{equation}\label{dPe_low energy}
dP^e=-\frac{2\alpha}{\pi\lambda_C}\frac{m_e}{\gamma}\Phi^\prime(0)\left(\frac{\chi_e}{\epsilon_\gamma}\right)^{2/3} d\epsilon_\gamma,~~~\epsilon_\gamma\ll 1,\chi_e. 
\end{equation} 
However the total probability of emission remains finite and the intensity of the radiation emission scales as $\epsilon_\gamma^{1/3}$ as $\epsilon_\gamma\rightarrow 0$. 

\subsection{Total probabilities of multiphoton Compton and Breit-Wheeler processes: Radiation length.}

If we integrate the expression for the differential probability of a photon emission by an electron (\ref{dPe}), we obtain the total probability of this process \cite{RitusJETP1964}: 
\begin{equation}
P^{e}(\chi_e)=-\frac{\alpha}{2\pi\lambda_C}\frac{m_e}{\gamma}\chi_e\int\limits_0^\infty dx\frac{5+7z+5z^2}{\sqrt{x}(1+z)^3}\Phi^\prime(x),~~~z=\chi_e x^{3/2}
\end{equation}
In the limiting cases of large and small $\chi_e$ it is possible to carry out the integration, and the total probability can be written down in the form of series in $\chi_e$:
\begin{equation}
P^{e}(\chi_e)=\left\{
\begin{tabular}{l}
$\displaystyle\frac{5}{3^{1/2}\pi}\frac{\alpha}{\lambda_C}\left(\frac{I}{I_S}\right)^{1/2}\left(1-\frac{8\sqrt{3}}{15}\chi_e+...\right),~~~\chi_e\ll 1$ \\ \\
$\displaystyle\frac{28\Gamma(2/3)}{9\pi }\frac{\alpha}{\lambda_C}\left(\frac{I}{I_S}\right)^{1/2}(3\chi_e)^{-1/3}\left(1-\frac{45}{28\Gamma(2/3)}\left(3\chi_e\right)^{-2/3}\right),~~~\chi_e\gg 1.$
\end{tabular}
\right.
\end{equation}
Here we explicitly show the dependence of the total probability on the EM field intensity, and $\Gamma(z)=\int_0^\infty t^{z-1}e^{-t}dt$ is the Euler-Gamma function. The same can be done in order to  obtain the electron-positron pair production probability \cite{RitusJETP1964}:
\begin{equation}\label{gamma probability}
P^{\gamma}(\chi_\gamma)=-\frac{\alpha}{32\pi\lambda_C}\frac{m_e}{\omega_\gamma}\chi_\gamma\int\limits_{(4/\chi_\gamma)^{2/3}}^\infty dx\frac{8z+1}{\sqrt{x}z\sqrt{z(z-1)}}\Phi^\prime(x),~~~z=\frac{\chi_\gamma x^{3/2}}{4}.
\end{equation}
In the limiting cases of large and small $\chi_\gamma$ it is possible to carry out the integration, and the total probability can be written down in the form of series in $\chi_\gamma$, the first term of which is given by
\begin{equation}\label{gamma probability limits}
P^{\gamma}(\chi_\gamma)=\left\{
\begin{tabular}{l}
$\displaystyle\frac{3\sqrt{3}}{16\pi\sqrt{2}}\frac{\alpha}{\lambda_C}\left(\frac{I}{I_S}\right)^{1/2}\exp\left(-\frac{8}{3\chi_\gamma}\right),~~~\chi_\gamma\ll 1,$ \\ \\
$\displaystyle\frac{30\cdot 2^{1/3}}{7\Gamma^2(1/6)}\frac{\alpha}{\lambda_C}\left(\frac{I}{I_S}\right)^{1/2}\left(3\chi_\gamma\right)^{-1/3},~~~\chi_\gamma \gg 1.$
\end{tabular}\right.
\end{equation}

\begin{figure}[tbp]
\epsfxsize10cm\epsffile{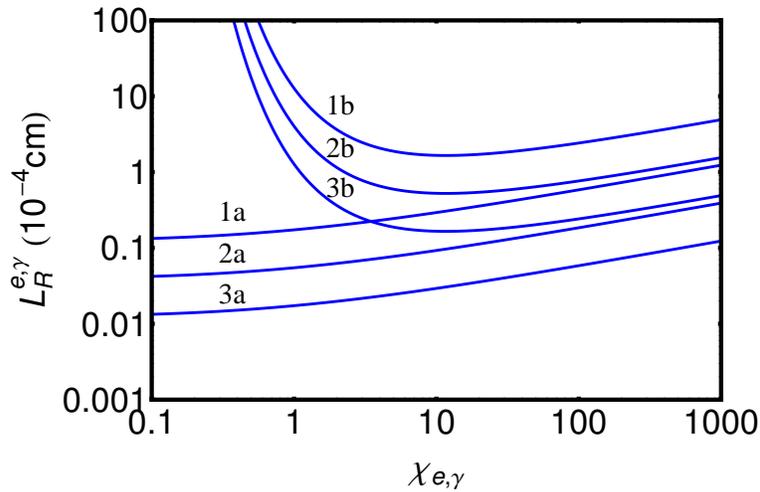}
\caption{(Color online) The dependencies of the electron/positron (a) and photon (b) mean free paths with respect to radiation in constant crossed EM field on parameter $\chi_{e,\gamma}$ for different intensities of the EM field: $10^{23}$ (1a, 1b), $10^{24}$ (2a, 2b), $10^{25}$ W/cm$^2$ (3a, 3b).}\label{L}
\end{figure}

The lifetime of an electron with respect to radiation, or a photon with respect to pair production, is $\tau_{e,\gamma}=\left(P^{e,\gamma}\right)^{-1}$. We can also define the radiation length, a mean free path with respect to either radiation or pair production, $\mathcal{L}_R^{e,\gamma}=\tau_{e,\gamma}$. For example, a 10 GeV electron interacting with a laser pulse with intensity of about $I\sim 10^{22}$ W/cm$^2$, such that $\chi_e\sim 1$, has the radiation length of about $\lambda_0/5$. At the same time a 10 GeV photon interacting with the same pulse ($I\sim 10^{22}$ W/cm$^2$) has a radiation length of about $10\lambda_0$. The fact that electron radiation length is fifty times smaller then that of a photon is connected with the huge enhancement of the probability for an electron to emit a low energy photon that can be seen from Eq. (\ref{dPe_low energy}) (also see Fig. \ref{dP}). The dependencies of the radiation lengths on parameter $\chi_{e,\gamma}$ are shown in Fig. \ref{L} for different intensities of the laser radiation. At high values of $\chi_{e,\gamma}$ both electron and photon radiation lengths increase as $(\chi_{e,\gamma})^{1/3}$. However at small values of $\chi_{e,\gamma}$ the behavior of $\mathcal{L}_R^{e}$ and $\mathcal{L}_R^{\gamma}$ is completely different. The electron radiation length tends to a constant value, while the photon radiation length tends to infinity. Such behavior is a good illustration of the fact that the process of pair creation does not have a classical analogy.       

We note that the photon radiation length increases for both small and large values of parameter $\chi_\gamma$. This means that there exists a value of $\chi_\gamma$ that minimizes the photon radiation length. From Eq. (\ref{gamma probability}) we find the value is $\chi_\gamma^{min}\approx 12$ and the minimal radiation length is
\begin{equation}
\mathcal{L}^\gamma_{R,min}\simeq\frac{5 \lambda_C}{\alpha}\left(\frac{I_S}{I}\right)^{1/2}=\frac{106\lambda_0}{a},
\end{equation} 
and it is inversely proportional to the square root of intensity. Here $\lambda_0=1$ $\mu$m. In order to maximize the number of events of a photon decaying into an electron/positron pair let's assume that $L^\gamma_{min}=\lambda_0$ and derive the intensity of the laser and the energy of photons. Then $a=106$, corresponding to the intensity of about $10^{22}$ W/cm$^2$. However the requirement of $\chi_\gamma^{max}\approx 12$ fixes the energy of photons: $\omega_\gamma\approx 12$ GeV. This configuration would require 12 GeV photons to be generated in abundance in the multiphoton Compton process. Further reduction of the radiation length is possible for higher intensities, however in this case, the 1D approximation for an electron, positron, and photon interaction with a laser pulse will no longer be valid, since such interaction will be greatly affected by the transverse dynamics of charged particle moving inside the laser pulse. Thus we can conclude here that the prolific pair production in the e-beam interaction with the intense laser pulse is closely connected with the strong transverse dynamics of electron and positrons as could have been expected from the results of Refs. \cite{BellKirk, FedotovPRL2010,BulanovPRL2010}.

\section{Electromagnetic cascade-type process.}

Using the results of the preceding section we can approximately model the interaction of an energetic electron beam with the intense circularly polarized laser pulse. We use the 1D approximation and the laser pulse is chosen to have a Gaussian profile. Since the radiation formation length, $\mathcal{L}_R$, is much smaller than the laser wavelength, we can adopt the locally constant field approximation. In this approximation the differential probabilities of the processes of a photon emission by an electron/positron and a photon decay into an electron-positron pair are calculated for the case of constant crossed EM field at each point, then the total probabilities or the differential ones are obtained by integrating over time and space. 

If we consider the evolution of the energy distributions of the electron, positron and photon beams ($f_{e^\pm}(\epsilon_{e^\pm}^\prime,t)$, $f_{\gamma}(\epsilon_{\gamma},t)$) inside the laser pulse, then at each time instant these distributions are sums of three terms. For an electron (positron) distribution these three terms are (i) the distribution of electrons (positrons) that did not emit a photon, (ii) the distribution of electrons (positrons) that emitted a photon, and (iii) the distribution of electrons (positrons) that were created as a result of a photon decay. For a photon distribution these three terms are (i) the distribution of photons that did not decay into an electron positron pair, (ii) and (iii) the distribution of photons that were emitted either by electrons or positrons. In other words the distribution functions obey the following recursive expressions: 
\begin{equation}
f_{e^\pm}(\epsilon_{e^\pm}^\prime,t+\Delta t)=f_{e^\pm}(\epsilon_{e^\pm}^\prime,t)\left[1-
P^{e}(\epsilon_{e^\pm}^\prime,t)\Delta t\right]
+\left\{\int\limits_0^1\left[f_{e^\pm}(x,t)\mathcal{P}_1(x,\epsilon_e^\prime,t)+f_\gamma(x,t)\mathcal{P}_2(x,\epsilon_e^\prime,t)dx\right]\right\}\Delta t, 
\end{equation}
\begin{equation}
f_{\gamma}(\epsilon_{\gamma},t+\Delta t)=f_{\gamma}(\epsilon_{\gamma},t)\left[1-P^{\gamma}(\epsilon_\gamma,t)\Delta t\right] 
+\left\{\int\limits_0^1 \left[f_{e^+}(x,t)+f_{e^-}(x,t)\right]\mathcal{P}_3(x,\epsilon_\gamma,t)dx\right\}\Delta t, 
\end{equation}
where $\mathcal{P}_1(\epsilon_e,\epsilon_e^\prime,t)=dP^{e}/d\epsilon_e^\prime$, $\mathcal{P}_2(\epsilon_\gamma,\epsilon_e^\prime,t)=dP^{\gamma}/d\epsilon_e^\prime$, and $\mathcal{P}_3(\epsilon_e,\epsilon_\gamma^\prime,t)=dP^{e}/d\epsilon_\gamma^\prime$ are the differential probabilities of the corresponding processes in a constant crossed EM field defined in the previous section. Correspondingly the functions $P^{e}(\epsilon_e,t)$ and $P^{\gamma}(\epsilon_\gamma,t)$ are the probabilities of a photon emission by an electron/positron, and a photon decay into electron-positron pair. It is emphasized here that they are the functions of initial electron/positron/photon energy and the instantaneous (at time $t$) value of the EM field. These expressions can be rewritten in a form of a system of differential equations for the distribution functions $f_i$, where $i=e^+,e^-,\gamma$, by dividing both sides by $\Delta t$ and then by taking limit at $\Delta t\rightarrow 0$:
\begin{equation}\label{kinetic1}
\frac{df_{e^\pm}(\epsilon_{e^\pm}^\prime,t)}{dt}=-f_{e^\pm}(\epsilon_{e^\pm}^\prime,t)P^{e}(\epsilon_{e^\pm}^\prime,t) +\int\limits_0^1\left[f_{e^\pm}(x,t)\mathcal{P}_1(x,\epsilon_e^\prime,t)+f_\gamma(x,t)\mathcal{P}_2(x,\epsilon_e^\prime,t)dx\right], 
\end{equation}
\begin{equation}\label{kinetic2}
\frac{df_{\gamma}(\epsilon_{\gamma},t)}{dt}=-f_\gamma(\epsilon_\gamma,t)P^{\gamma}(\epsilon_\gamma,t)+\int\limits_0^1 \left[f_{e^+}(x,t)+f_{e^-}(x,t)\right]\mathcal{P}_3(x,\epsilon_\gamma,t)dx.
\end{equation}
These equations are analogous to the ones obtained in Refs. \cite{kinetic,Sokolov_kinetic,kinetic_Bell}

\begin{figure}[tbp]
\epsfxsize6cm\epsffile{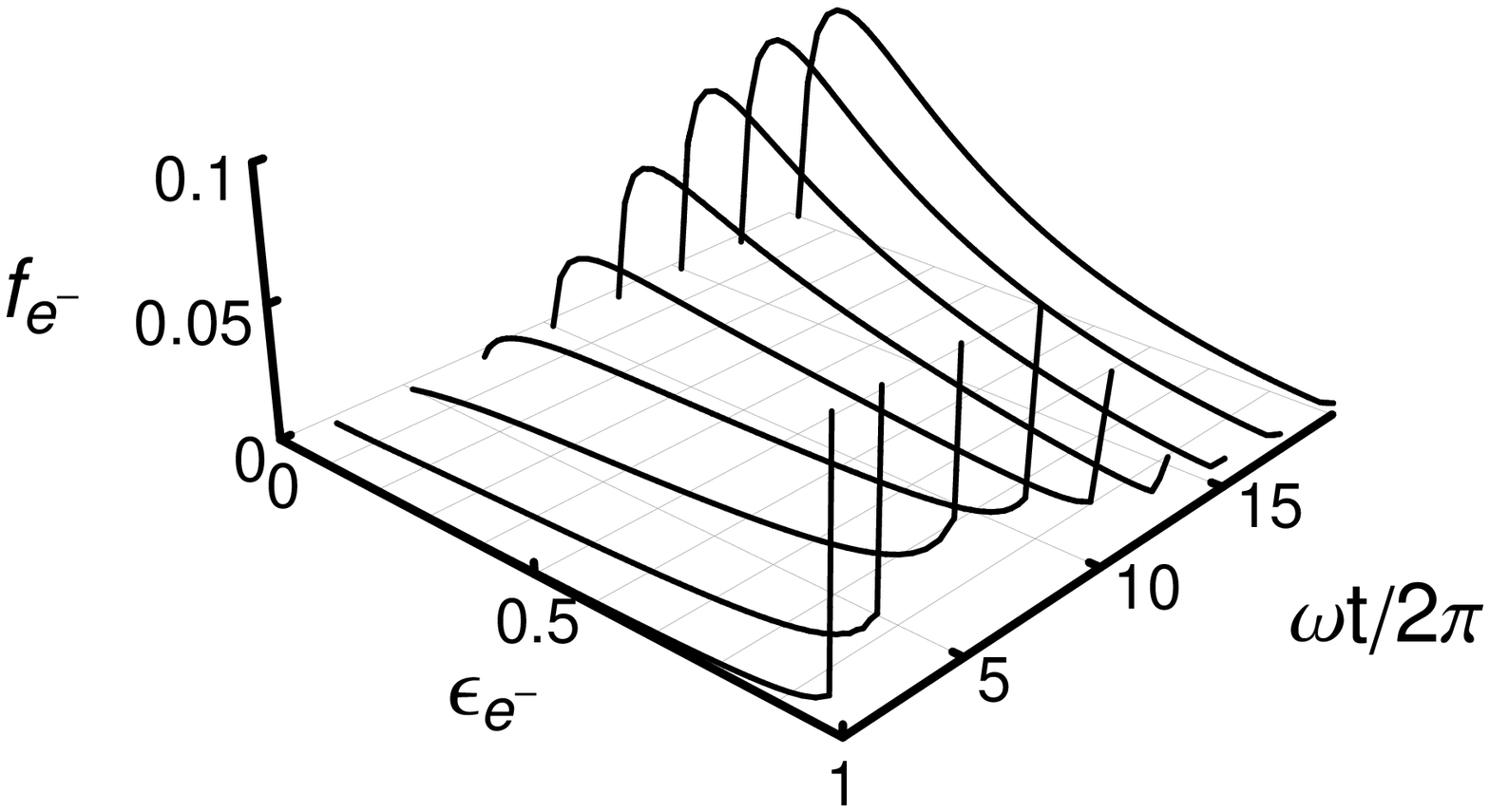}\epsfxsize6cm\epsffile{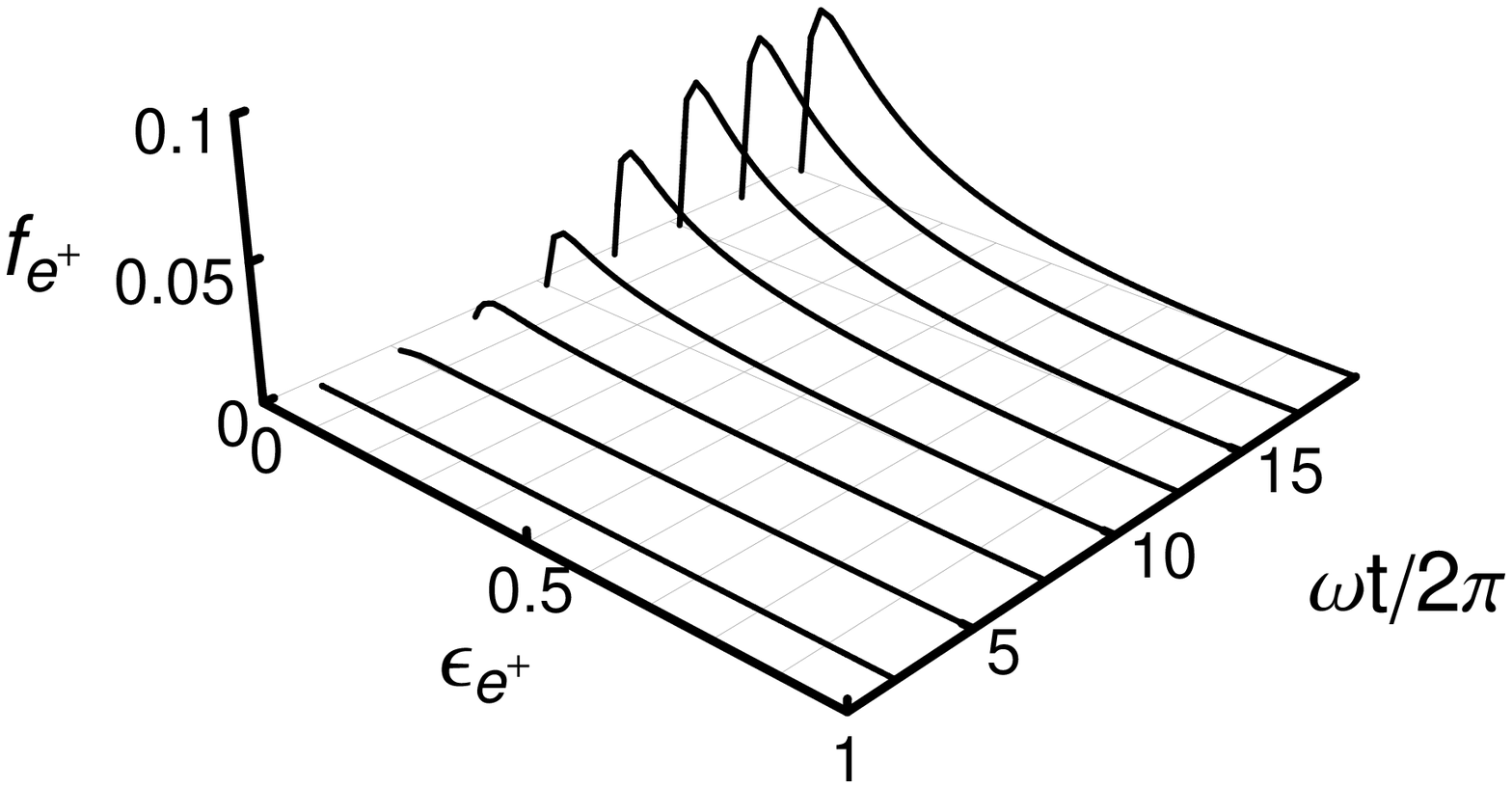}\epsfxsize6cm\epsffile{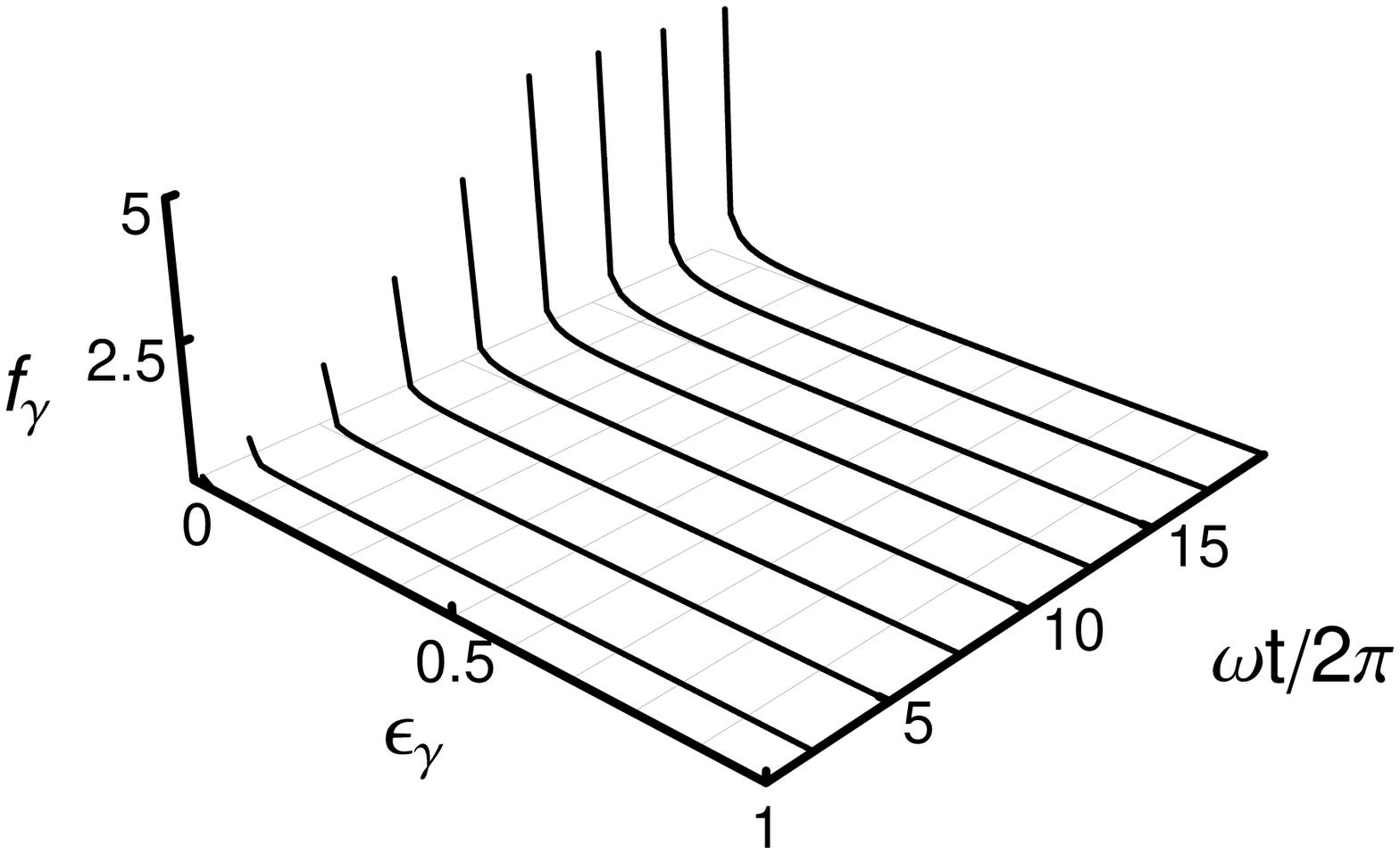}
\caption{The evolution of the spectra of electrons, positrons and photons during the interaction of a 10 GeV electron beam with the $2.5\times 10^{22}$ W/cm$^2$ laser pulse. The duration of the laser pulse is ten waveperiods. The intensity profile was chosen to be Gaussian with duration of 30 fs.}\label{evolution}
\end{figure}

This system of equations, (\ref{kinetic1}) and (\ref{kinetic2}), can be solved numerically. The results are presented in Fig. \ref{evolution}, where the evolution of the electron, positron and photon spectra are shown during the interaction of the beam and the pulse. The initial electron beam was chosen to be monoenergetic with the energy of 10 GeV, the laser field has Gaussian profile with the duration of ten waveperiods. The peak intensity is $2.5\times 10^{22}$ W/cm$^2$. The electrons undergo a cascade-type process involving the emission of multiple photons, which in their turn can decay into electron-positron pair, giving rise to the positron beam. Such evolution of the electron beam leads to a fast loss of the beam energy, which can be seen from Fig. \ref{evolution}a. Being initially monoenergetic the electron beam transforms into a broad distribution with the maximum near 1 GeV, losing almost 70\% of its initial energy. Moreover such form of the spectrum leads to the reduction of the emission of high energy photons, which can be seen from Fig. \ref{evolution}c, where the photon spectrum evolution is shown. This spectrum demonstrates an exponentially decaying form, i.e. the number of high energy photons is highly suppressed when compared to the low energy ones. The number of emitted photons is about twelve times higher then the number of initial electrons, \textit{i.e.}, on average each electron undergoes the photon emission twelve times.   

\begin{figure}[tbp]
\epsfxsize6cm\epsffile{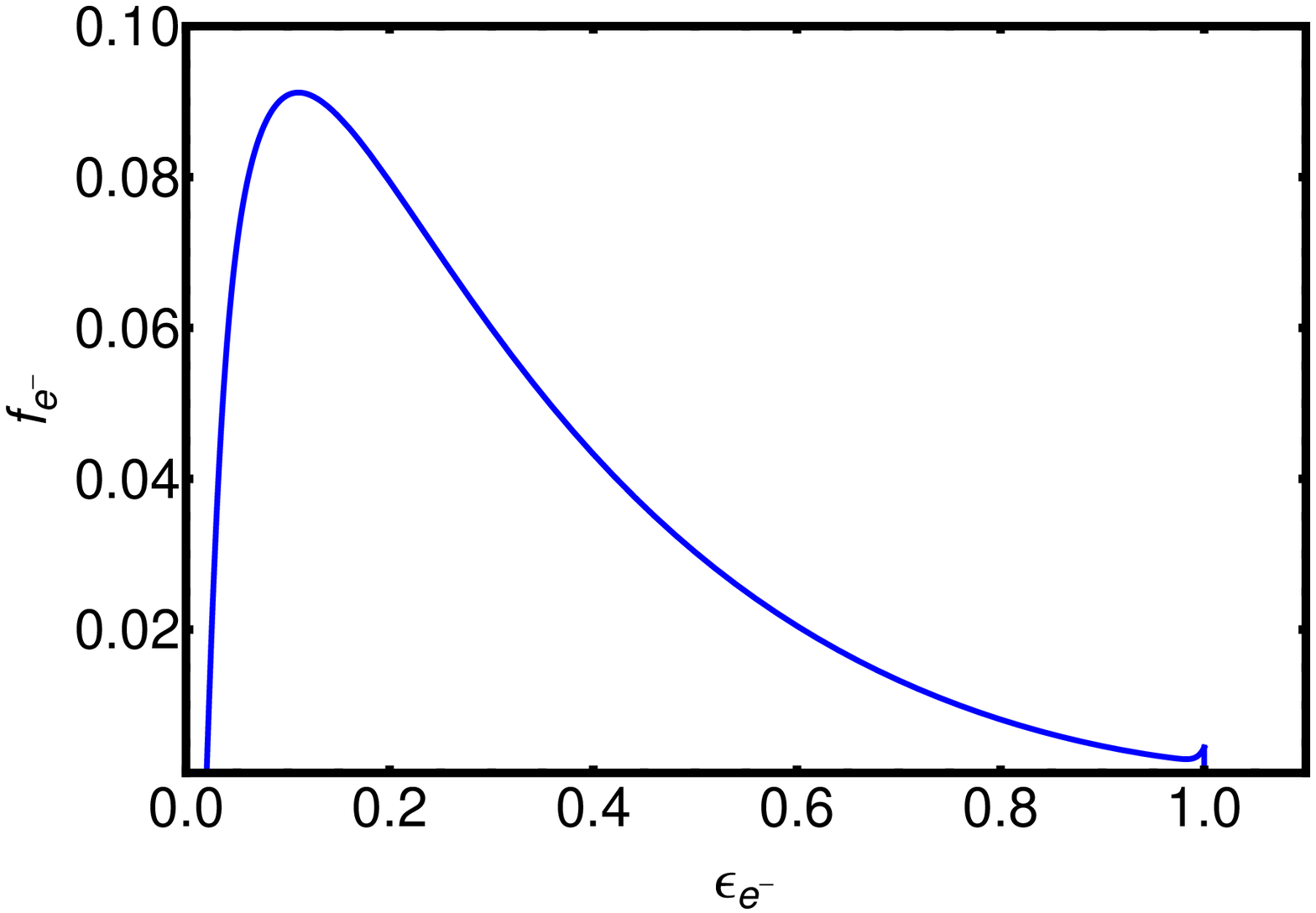}\epsfxsize6cm\epsffile{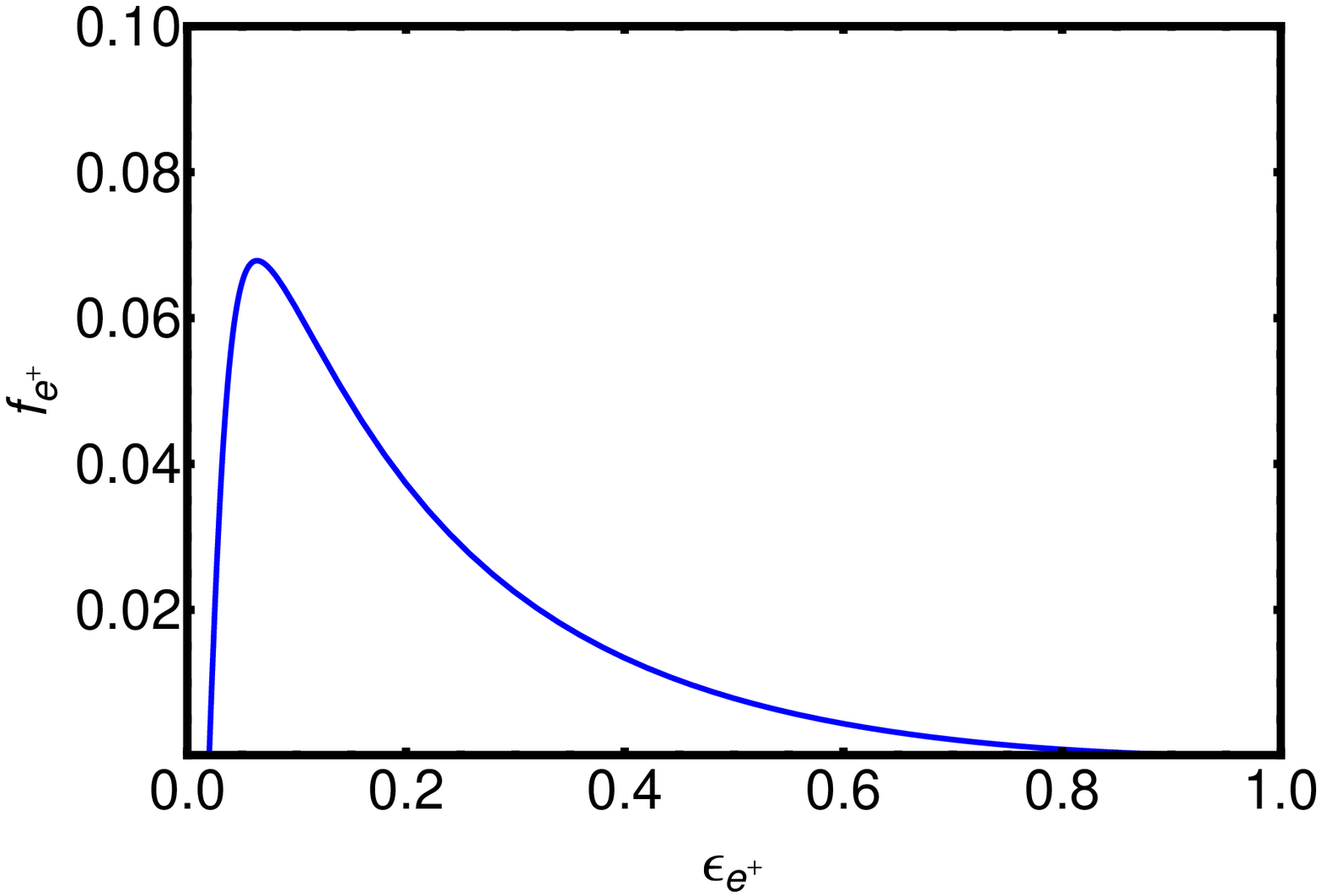}\epsfxsize6cm\epsffile{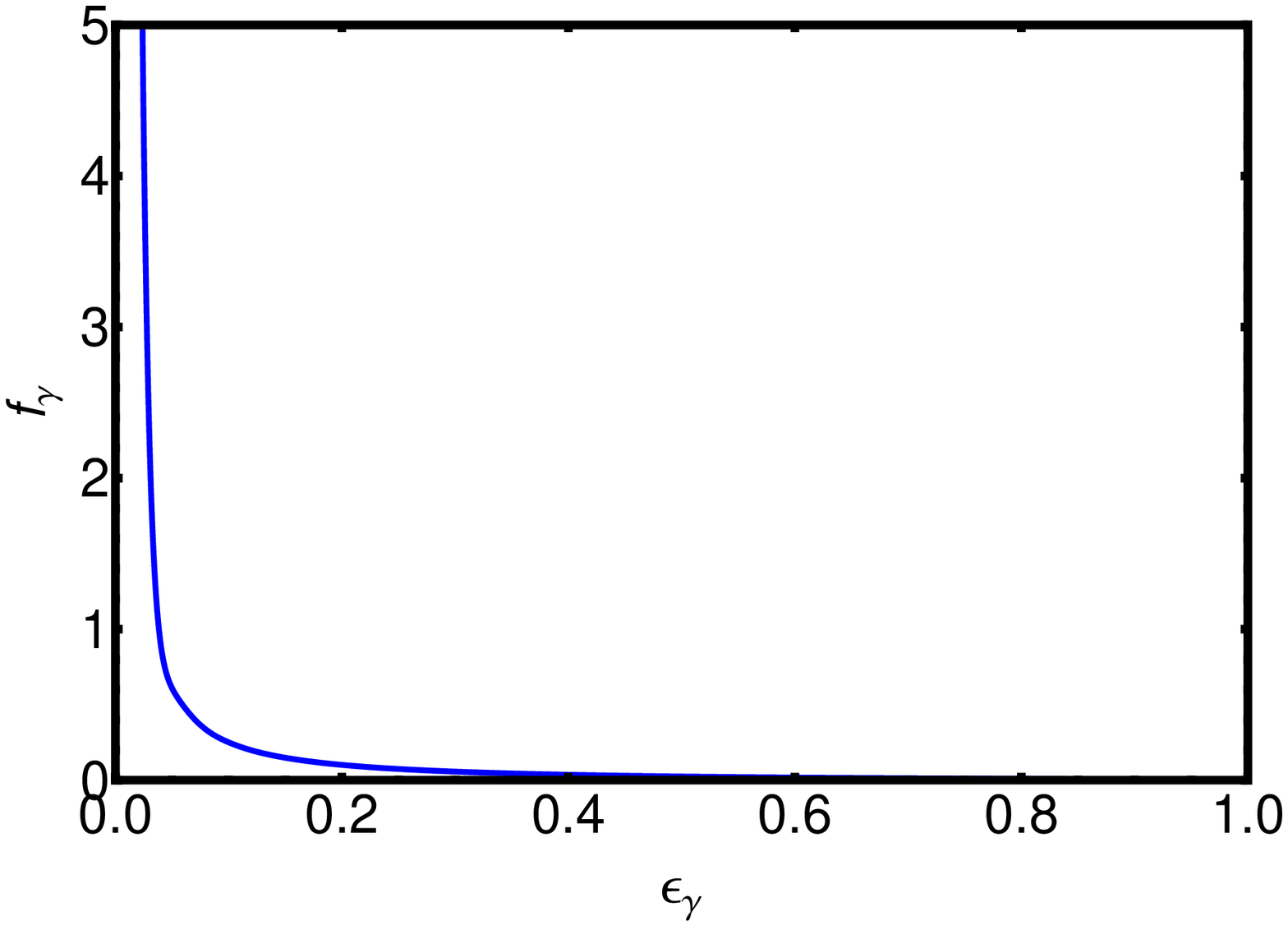}
\caption{(Color online) The spectra of electrons, positrons and photons after the interaction of a 10 GeV electron beam with a $2.5\times 10^{22}$ W/cm$^2$ laser pulse. The duration of the laser pulse is ten waveperiods. The intensity profile was chosen to be gaussian.}\label{final spectra}
\end{figure}

As discussed in the previous section, the decay of an energetic photon into an electron-positron pair in the presence of a strong EM field leads to a spectrum of electrons/positrons, the form of which depends on the value of parameter $\chi_\gamma$. If $\chi_\gamma>8$ then there are two peaks in the spectrum, corresponding to $\epsilon_e\sim 1-4/3\chi_\gamma$ and $\epsilon_e\sim 4/3\chi_\gamma$. However the spectrum of photons is dominated by the low energy ones and the contribution of the $\gamma\rightarrow ee$ decays with the two-peak structure is negligible. Most electron-positron pairs are generated by the photons with $\chi_\gamma<8$. It is plausible to assume that the main contribution to the positron spectrum comes from the photons, whose radiation length is about the length of the laser pulse ($10\lambda$). The parameter $\chi_\gamma$ corresponding to such radiation length ($\mathcal{L}_R^\gamma\simeq 10\lambda$) and laser intensity of $2.5\times 10^{22}$ W/cm$^2$ is $\chi_\gamma\simeq 1.5$. For such photons the most probable is the decay into electron-positron pair, where the final energy is equally distributed between the electron and the positron, \textit{i.e.}, $\epsilon_{e^+}=\epsilon_{e^-}\simeq 800$ MeV. This estimate is in a good agreement with the result of numerical solution of Eqs. (\ref{kinetic1}) and (\ref{kinetic2}), which show a maximum in positron spectrum at  $\epsilon_{e^+}\simeq 640$ MeV, if we take into account the energy loss of positrons due to the photon emission in the laser pulse. The number of produced positrons is almost the same as the number of initial electrons. The final forms of the spectra, when the interaction of the beam and the laser pulse is over, are shown in Fig. \ref{final spectra}. 

\subsection{Comparison with the solution of classical equations of motion in the presence of radiation reaction}

In what follows we consider the equation of motion of an electron in the EM field taking into account the radiation reaction in order to compare the results obtained in the framework of nonlinear QED with those of the classical electrodynamics:
\begin{equation} \label{eq of motion}
m_e\frac{du^\mu}{ds}=eF^{\mu\nu}u_\nu+g^\mu,
\end{equation}   
Here $u_\mu=(\gamma,\mathbf{p}/m)$ in the four-velocity and $s=\int dt/\gamma$ is the proper time. The radiation friction force is taken in the Landau-Lifshitz form \cite{L-L} in order to avoid unphysical self-accelerating solutions, which are possible if the radiation friction force is taken in the Lorentz-Abraham-Dirac (LAD) form \cite{LAD} (for the discussion of the equivalence of the LAD and L-L forms of the radiation friction force see \cite{LAD vs L-L}). The force $g^\mu$ is defined in such a way that an integral of it performed over the world line of an electron moving in the EM field, is equal to the total emitted photon momentum with negative sign. 

We next consider a model case of a 1D motion of an electron in the EM field of a counterpropagating laser pulse in order to be able to compare the results of classical calculations with the quantum calculations described above. If $\epsilon_{rad}a\gamma_e^2\gg 1$, then the interaction is purely dissipative and we can neglect all the EM forces except the radiation reaction. In this limiting case the equation of motion is reduced to \cite{LAD vs L-L}
\begin{equation}\label{RR}
\frac{dp}{dt}=-\epsilon_{rad}\omega_0a^2(-2t)\frac{p^2}{m}.
\end{equation} 
Since we assumed a 1D motion of an electron in the EM field of a counterpropagating laser pulse, then $x\approx -t$ and $a(x-t)\approx a(-2t)$. The solution of the equation for the longitudinal momentum component is 
\begin{equation}
p(t)=-\frac{p_0}{1+\epsilon_{rad}\omega_0 (p_0/m_e)\int^t_0a^2(-2\eta)d\eta},
\end{equation}
which shows gradual decrease of the electron momentum as it passes through the laser pulse due to radiation. Here $p_0$ is the initial electron momentum.

It is well known that in the classical approximation the amount of the radiation emitted by an electron moving in the EM field is overestimated. Thus the electron energy is more rapidly depleted. It is connected with the fact that classical formula for the radiation intensity allows for the emission of photons with the energy greater than the initial electron energy. Eq. (\ref{eq of motion}) can be modified in a way that the integral of the radiation friction force performed over the electron world line will be equal to the total emitted momentum, but the emitted momentum is calculated in the QED framework. From Eq. (\ref{dPe}) we can get the expression for the radiation intensity \cite{RitusJETP1964}:
\begin{equation}
I=I_{cl}G(\chi_e),~~~\mbox{where}~~~G(\chi_e)=1-\frac{55\sqrt{3}}{16}\chi_e+48\chi_e^2+...,~~~\chi_e\ll 1,
\end{equation} 
here $I_{cl}=2e^2m_e^2\chi_e^2/3$ is the classical radiation intensity. Then, following \cite{BellKirk}, we can modify the classical equation of motion by multiplying the radiation friction force by the function $G(\chi_e)$, $g_\mu^Q=g_\mu G(\chi_e)$, in order to reduce the amount of unphysical energy loss by the electron due to the overestimation of emitted radiation.

\begin{figure}[tbp]
\epsfxsize6cm\epsffile{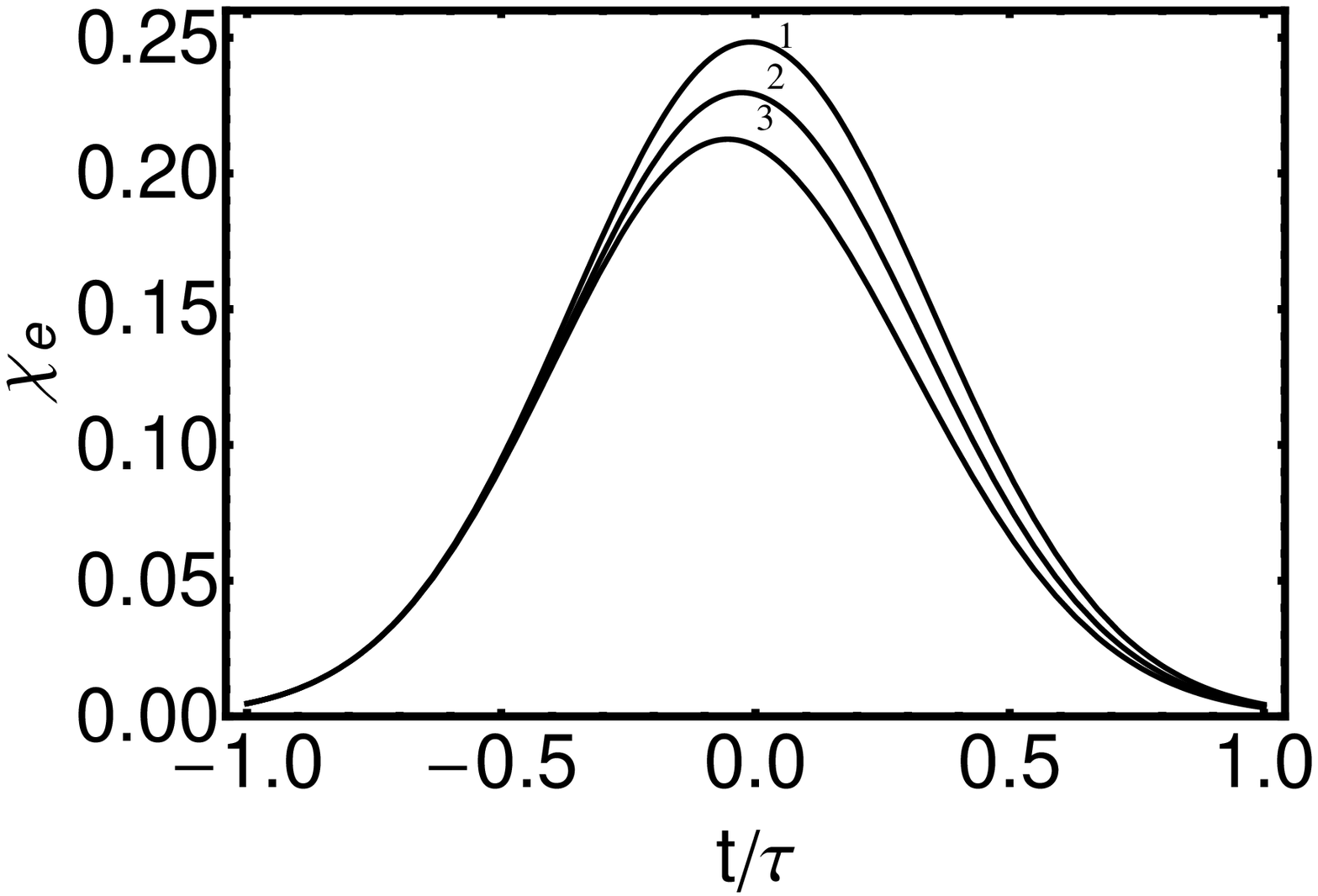}\epsfxsize6cm\epsffile{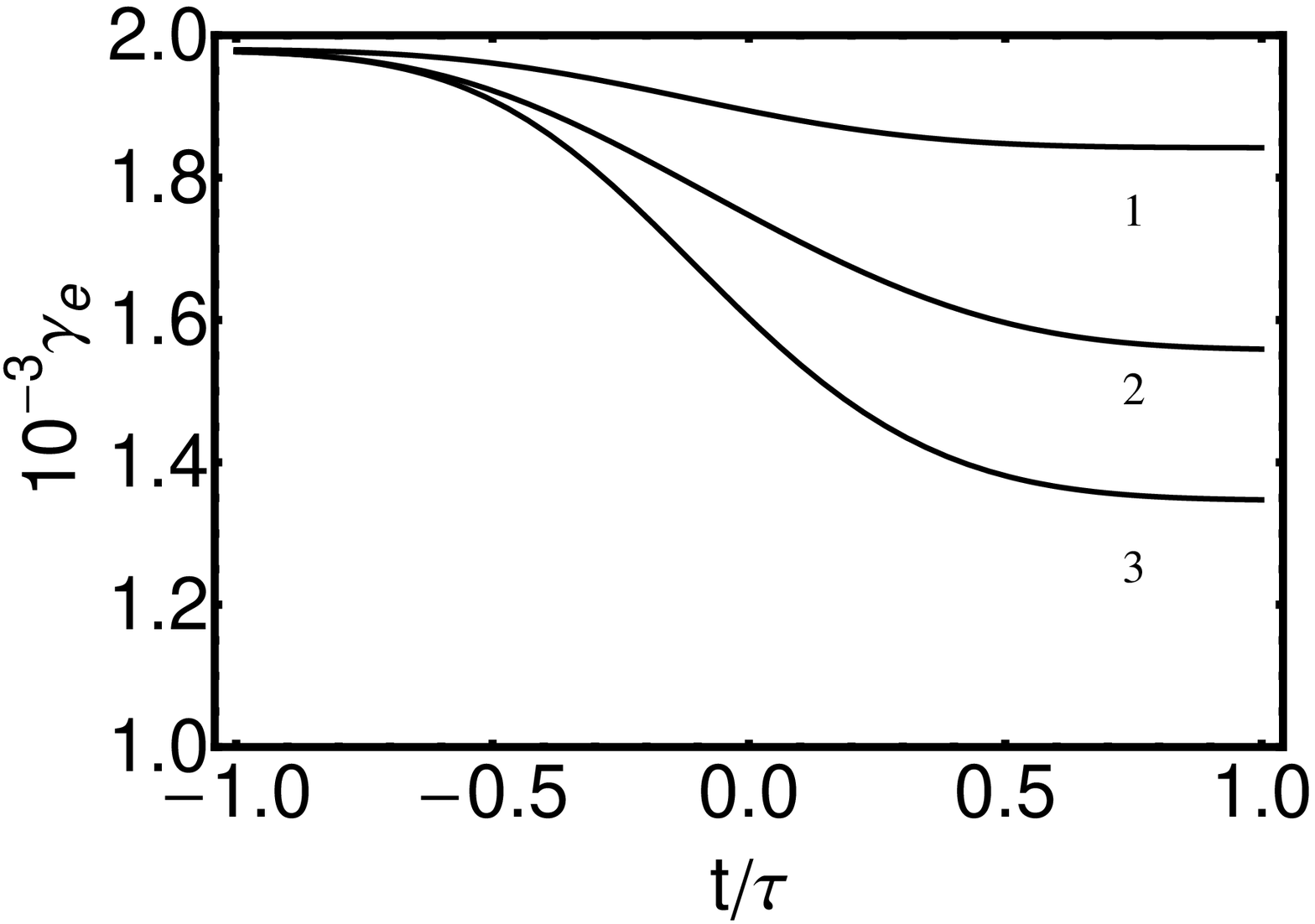}\epsfxsize6cm\epsffile{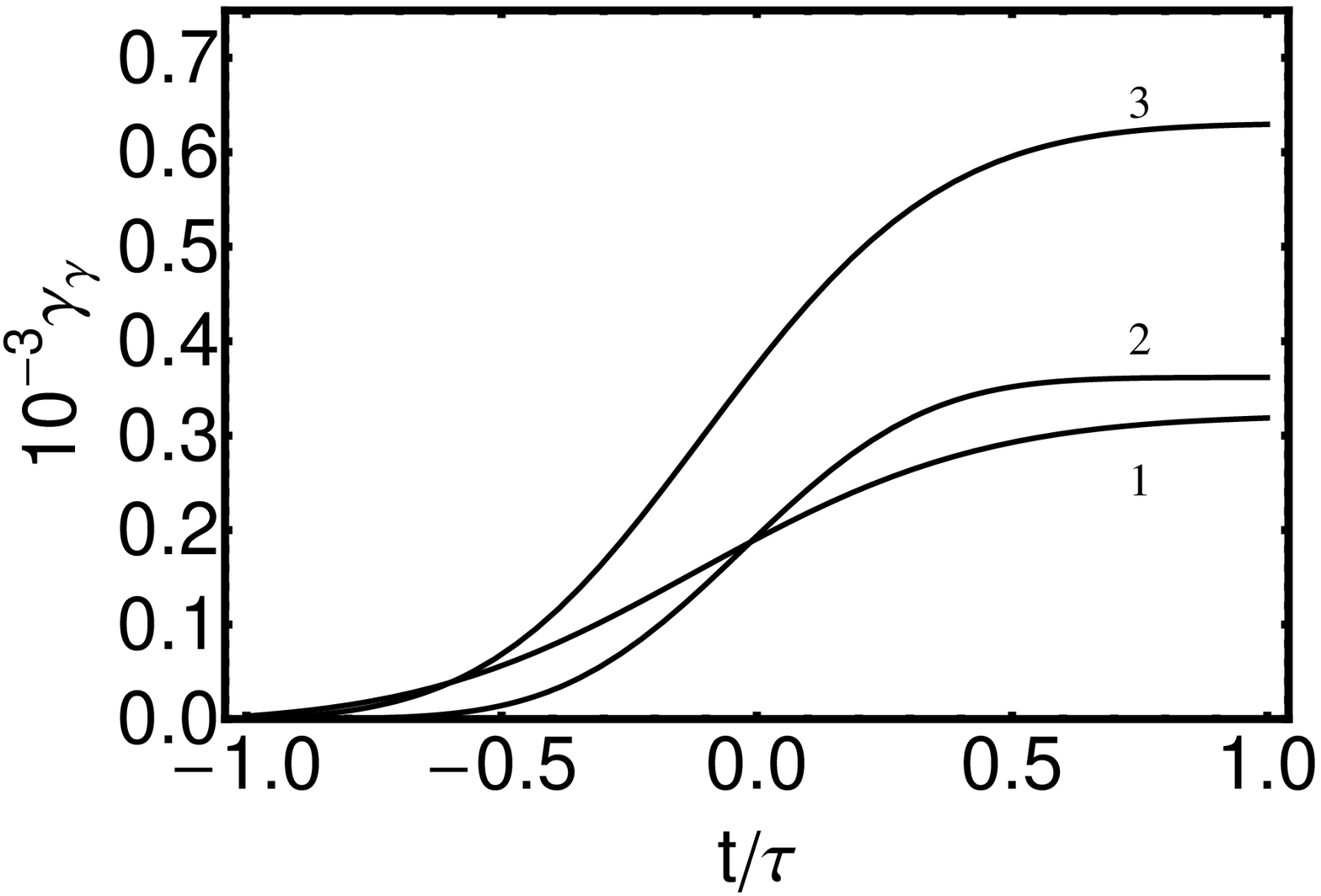}
\caption{The dependence of a) the parameter $\chi_e$, b) the electron energy, and c) the radiated intensity on time for the 1 GeV electron beam interaction with a $10^{21}$ W/cm$^2$ laser pulse with a duration of 10 cycles. (1) - "Quantum" corresponds to the results obtained through the solution of kinetic equation (\ref{kinetic1},\ref{kinetic2}), (2) - "modified classical" corresponds to the results obtained through the solution of the modified classical equation of motion, and (3) - "classical" corresponds to the results obtained through the solution of the classical equation of motion}. \label{comparison}
\end{figure}  

We now compare the evolution of the e-beam energy and intensity of emitted radiation in three cases: (i) "quantum", according to the solution of kinetic equations (\ref{kinetic1},\ref{kinetic2}); (ii) "modified classical", according to the solution of (\ref{eq of motion}) with the modified radiation friction force, $g_\mu^Q$; (iii) "classical", according to the solution of Eq. (\ref{eq of motion}). We consider the case of a 1 GeV electron beam interacting with a $10^{21}$ W/cm$^2$ gaussian laser pulse with a duration of ten cycles. The laser wavelength is 1 $\mu$m. In Fig. \ref{comparison} we present the dependencies of parameter $\chi_e$, e-beam energy, and radiated intensity on time. In the "quantum" case the energy, which is plotted, is the average energy of the beam. The evolution of parameter $\chi_e$ in each of the three cases is shown in figure 4a. In the quantum case the e-beam is able to reach the peak intensity of the laser pulse with highest energy, which is resulting in highest values of $\chi_e$. The evolution of the e-beam energy is shown in Fig. \ref{comparison}b. One can see a significant difference between final e-beam energy in these three cases. As was already mentioned, the classical approach overestimates the amount of the radiated energy and thus the electron looses more energy. However in the "modified classical" approach this discrepancy is approximately accounted for (see Fig. \ref{comparison}c where the evolution of the radiated energy is shown). Nevertheless the "modified classical" and quantum cases give different values of the final e-beam energy. It is connected with the fact that in both "classical" and "modified classical" cases the electron energy loss is due to emission and the total emitted energy is determined only by the electron energy loss. In the "quantum" case the energy balance is determined by the energy-momentum conservation (\ref{energy conservation}), where the absorption of multiple photons from the background EM field is present. As we can see from the results presented in Fig. \ref{comparison} the total amount of energy absorbed from the laser pulse accounts for almost half of the emitted energy. Thus the "quantum" approach is not only giving the correct amount of radiated energy, it also shows that the e-beam energy loss is not as severe as it could have been expected from "classical" considerations, and that taking into account the energy absorption from the laser pulse is crucial for the understanding of the dynamics of energetic e-beam interaction with the laser pulse.          

\section{Conclusions.} 

We considered the interaction of a high energy electron beam and an intense laser pulse with the electrons and positrons undergoing the multiphoton Compton process, and with the emitted photons undergoing the multiphoton Breit-Wheeler process. The parameters of interaction were chosen in a way that the transverse dynamics of the electron beam (and the emerging positron beam from the interaction) could be neglected, \textit{i.e.}, electron beam energy $\sim 10$ GeV and laser pulse peak intensity $10^{21-23}$ W/cm$^2$. This implies that a transverse quiver amplitude of the electron, $\sim \lambda_0 a/\gamma $, is much less than the laser spot size $r_0$. It is known that the individual processes of a photon emission by an electron or positron and a photon decay into an electron-positron pair in a strong EM field exhibit an enhancement in the number of events producing the high energy photons or electrons/positrons, \textit{i.e.}, the processes exhibit highly asymmetric distributions of energy among the final state particles. One can expect that such enhancement, in principle, could lead to prolific electron-positron pair production, which is an important step in inciting the EM avalanche-type discharge. From the analysis of the mean free paths of electrons/positrons with respect to radiation, and photons with respect to pair creation, we concluded that in order to have efficient pair creation one should minimize the photon mean free path. However, this would require the utilization of extremely high intensity laser pulses, for which the  approximations assumed  in this paper are no longer valid and the electron/positron dynamics is highly affected by the transverse motion induced by EM fields of the laser. Thus the prolific pair creation is closely connected with the strong transverse dynamics of electrons/positrons in the EM field.

In order to investigate further the possibility of the prolific electron-positron pair production we solved the system of equations for the distribution functions of electrons, positrons, and photons in the strong electromagnetic field, describing the collision of an electron beam with the laser pulse. We showed that in this case the enhancement is suppressed. It is due to the fact that, inside the laser pulse, electrons undergo a cascade-type process involving multiple emissions of photons, leading to a fast depletion of the electron beam energy as well as the transformation of the electron spectrum from initially monoenergetic to a broad one with the maximum at low energies and decaying according to power law high energy tail. The spectrum of photons demonstrate a power law dependence with increased production of low energy photons. Some fraction of the photons convert into electron-positron pairs giving rise to the positron beam. The spectrum of positrons is similar to that of the electrons: broad with maximum at low energies and decaying high energy tail. The main contribution to the positron spectrum comes from the photons, whose mean free path is about the laser length. It is due to the fact that lower energy photons have longer mean free paths and escape the laser pulse without decaying into an electron-positron pair, and the number of high energy photons is suppressed by the power law dependence of the spectrum. This conclusion is in agreement with the results of the numerical solution of the equations for the distribution functions.    

In order to study the energy depletion of the electron beam due to radiation we compared different approaches for the description of the e-beam interaction with the laser pulse. We considered the e-beam energy loss in the cases of "classical" (electron equation of motion with the radiation friction force), "modified classical" (friction force takes into account quantum expression for radiation intensity), and "quantum" approaches. The energy absorption from the laser pulse plays a crucial role in the description of the e-beam dynamics in the intense laser pulse, since it accounts for a significant part of the high frequency radiation emitted by electrons leading to a smaller reduction in e-beam energy than is predicted by the classical and "modified classical" equations. We conclude that the energy absorption from the laser pulse, which is only present in the "quantum" approach, plays an important role in the description of the high energy electron interaction with the intense laser pulse. Thus the energy loss of the e-beam will provide a clear experimental observable for the transition from the classical to quantum regime of interaction in the planned experiments on multi-GeV e-beam interaction with PW-class laser pulses \cite{BulanovAAC12,BulanovPRL2010}.      

We would like the thank M. Chen, T. Heinzl, A. Di Piazza, and J. Wurtele for discussions. We appreciate support from the NSF under Grant No. PHY-0935197, the Office of Science of the US DOE under Contract No. DE-AC02-05CH11231.

\end{document}